\documentclass[twocolumn]{aastex62}
\usepackage{amsmath,amssymb,pifont}
\usepackage[cal=boondox]{mathalfa}

\newcommand\gaia{{\em Gaia}}
\newcommand\gadget{{\sc P-Gadget3}}

\newcommand\nbody{{\sc Nbody6++}}
\graphicspath{{./}{figures/}}
\submitjournal{ApJL}
\shorttitle{Dating globular cluster streams with Gaia}
\shortauthors{Bose et al.}
\begin{document}

\title{DATING THE TIDAL DISRUPTION OF GLOBULAR CLUSTERS WITH GAIA DATA ON THEIR STELLAR STREAMS}

\correspondingauthor{Sownak Bose}
\email{sownak.bose@cfa.harvard.edu}

\author[0000-0002-0974-5266]{Sownak Bose}
\affil{Harvard-Smithsonian Center for Astrophysics, 60 Garden Street, Cambridge, MA 02138, USA}

\author{Idan Ginsburg}
\affil{Harvard-Smithsonian Center for Astrophysics, 60 Garden Street, Cambridge, MA 02138, USA}

\author{Abraham Loeb}
\affil{Harvard-Smithsonian Center for Astrophysics, 60 Garden Street, Cambridge, MA 02138, USA}

\begin{abstract}

The \gaia{} mission promises to deliver precision astrometry at an
unprecedented level, heralding a new era for discerning the kinematic
and spatial coordinates of stars in our Galaxy. Here, we present a new
technique for estimating the age of tidally disrupted globular cluster
streams using the proper motions and parallaxes of tracer stars. We
evolve the collisional dynamics of globular clusters within the
evolving potential of a Milky Way-like halo extracted from a
cosmological $\Lambda$CDM simulation and analyze the resultant streams
as they would be observed by \gaia{}. The simulations sample a variety
of globular cluster orbits, and account for stellar evolution and the
gravitational influence of the disk of the Milky Way. We show that a
characteristic timescale, obtained from the dispersion of the proper
motions and parallaxes of stars within the stream, is a good indicator
for the time elapsed since the stream has been freely expanding away
due to the tidal disruption of the globular cluster. This timescale,
in turn, places a lower limit on the age of the cluster. The age can
be deduced from astrometry using a modest number of stars, with the
error on this estimate depending on the proximity of the stream and
the number of tracer stars used.

\end{abstract}


\section{Introduction} \label{sec:intro}

Globular clusters (GCs) are dense, spherical, concentrations of stars
with a characteristic mass of $\sim 10^5\, M_{\odot}$ (see
\citealt{Brodie-Strader} for a review) and half-light radii of $r_{h}
\sim$ 3--10 pc \citep[e.g.][]{Jordan2005,vdBergh2008}. It is
universally agreed that GCs are among the oldest observable objects
and are found in numerous galaxies, with giant elliptical galaxies
harboring thousands of them \citep{Peng2011}. Surveys have discovered
about 150 GCs around the Milky Way
\citep[e.g.][]{Kharchenko2013}. Understanding the nature and origin of
these GCs has important implications for not only the formation
history of the Milky Way, but also for models of structure and galaxy
formation \citep[e.g.][]{KravtsovGnedin,MBK2017}.

GCs are grouped into two sub-populations depending on
metallicity. So-called ``blue clusters" are metal-poor whereas, ``red
clusters" are metal-rich
\citep{ZinnWest,Usher2012,Roediger2014}. \citet{Renaud2017} argued
that blue clusters form in satellite galaxies and are accreted onto
the Milky Way, whereas red clusters form {\it in situ}. Furthermore,
\citet{KunduW} noted that blue GCs appear to be $\sim$ 20\% larger
than their redder counterparts. However, it is unclear whether this is
a real physical phenomenon or a projection effect \citep{Larsen2003}.

GC ages are typically acquired from studies of color-magnitude
diagrams in conjunction with stellar evolution models
\citep{Forbes10,Correnti16,Powalka17,Kerber18}. Such studies show that
Galactic GCs (GGCs) typically have ages over 10 Gyr, often exceeding
12 Gyr. However, there are also young GGCs such as Whiting 1, which
has an estimated age of $\sim 6$ Gyr \citep{Carraro,Valcheva15}. As
they traverse the Milky Way's potential, GGCs evaporate over time and
leave extended tidal tails \citep{Gnedin97,Fall01,Carlberg}.

In this Letter we explore a novel technique to date the stream
associated with a tidally disrupted GC. This, in turn, can be used to
constrain the gravitational potential and the assembly history of the
Milky Way \citep[e.g.][]{Johnston1999,Pearson2015,Bonaca2018}. Our
method relies on measurements of positions and motions of stars in the
plane of the sky, which will become available soon with the \gaia{}
DR2 catalog. \gaia{} is expected to provide precise astrometric
measurements for over a billion stars in the Milky Way, with proper
motions and parallaxes determined to 1 percent or better up to $\sim$
15 kpc \citep{PancinoGG}.

The remainder of this Letter is organized as follows: in
Section~\ref{sec:numerical}, we discuss the numerical techniques used
and the parameter space that we explore. In Section~\ref{sec:gaia} we
visualize the simulated streams as they might be detected by \gaia{},
and in Section~\ref{sec:timescale}, we discuss our method for age
determination. Finally, we summarize our conclusions in
Section~\ref{sec:conclusions}.

\section{Numerical methods} \label{sec:numerical}

In the following subsections, we outline the details of our numerical
setup.

\subsection{The galactic potential of the Milky Way}

We consider the combined effect of the time-evolving potential of the
smooth dark matter halo component of a Milky Way-like galaxy, as well
as the contribution of a central disk galaxy. For the halo component,
we first create zoom-in initial conditions of a Milky Way-like halo
($M_{200} =1.5 \times 10^{12} \, M_\odot$) extracted from the dark
matter-only {\it Copernicus Complexio} simulations
\citep{Bose2016,Hellwing2016}. Here, $M_{200}$ is the mass contained
within $r_{200}$, the radius within which the mean density is 200
times the critical density of the Universe. The halo that we have
chosen for re-simulation exhibits a fairly typical accretion history
for halos of this mass, and was evolved from redshift $z=127$ to $z=0$
using the \gadget{} code \citep{Springel2008}.

At each simulation output, we compute the gravitational potential
associated with the high-resolution particle distribution, and find
that its evolution with redshift is captured well by an analytic
expression of the form \citep{Buist2014,Renaud2015}:
\begin{equation} \label{eq:potential}
\phi(r,z) = -\frac{G M_s(z)}{r} \ln \left( 1 + \frac{r}{r_s(z)} \right)\;,
\end{equation}
where $G$ is Newton's constant, $r_s(z)$ is the scale radius of the
halo, while $M_s(z)$ is the mass contained within this radius. The
redshift evolution of these parameters can be written in the form:
\begin{equation} \label{eq:evolve}
\begin{aligned}
M_s(z) &=& M_s(0) \exp(-0.08z), \\
r_s(z) &=& r_s(0) \exp(-0.05z),
\end{aligned}
\end{equation}
where $M_s(0) = 2 \times 10^{11}\,M_\odot$, $r_s(0) = 20$ kpc,
and the coefficients of the exponentials are obtained by fitting the
potential from snapshots of our re-simulation.

The contribution of a central disk galaxy is modeled using a
superposition of three Miyamoto-Nagai disk potentials
\citep{Miyamoto1975}, following the procedure described by
\cite{Smith2015}. The disk is assigned a total mass of $5 \times
10^{10} \, M_\odot$ with a scale length of 3 kpc and a scale height of
300 pc. The disk potential does not evolve dynamically.

\subsection{Globular cluster dynamics}

Initial conditions for the GC simulations were generated using the
{\sc McLuster} code \citep{Kuepper2011}. We sample the positions and
velocities of star particles in the GCs according to a King profile
\citep{King1966}. In each case, the total initial mass of the star
cluster is set to $10^5 \, M_\odot$, with a concentration parameter
$W_0=6$. In five out of the six GC simulations that we run, we set the
half-mass radius $r_h = 8$ pc, which is a relatively large value for
typical GCs. In a final simulation, therefore, we additionally include
an example with $r_h = 3$ pc. Masses of individual stars are assigned
by sampling a \cite{Salpeter1955}-like initial mass function
(IMF). For computational reasons, we truncate the low-mass end of the
IMF to $\sim 0.9 \, M_\odot$, resulting in a total of 21,730 star
particles in the $10^5\,M_\odot$ cluster.

The dynamical evolution of these star clusters is performed using the
publicly available \nbody{} code \citep{Wang2015}. \nbody{} is a
hybrid MPI-GPU accelerated version of the {\sc Nbody6} code
\citep{Aarseth2003,Nitadori2012}, and contains routines accounting for
the evolution of star clusters in background tidal fields developed as
part of {\sc Nbody6-tt} \citep{Renaud2015a}. The simulations (and,
therefore, the halo potential in Eq.~\ref{eq:potential}) are
initialized at $z=7$ using the same random seed and evolved for 12.96
Gyr, corresponding to the time interval until $z=0$. Mass loss through
stellar evolution is included in all of our simulations. The initial
configurations for all the simulations that we have performed are
summarized in Table~\ref{tab:sims}, where we have also listed the
orientations of the position/velocity vectors with which the GCs are
initialized. Note that the parameter space that we explore in our
simulations is not designed to be representative of the population of
GCs in the Milky Way.

\begin{table}
\centering
\caption{A summary of the numerical experiments performed. All
  distances and velocities are measured relative to the center of the
  galactic potential. The GC orbits, which for simplicity we have
  constrained to lie solely in the $x$-$z$ plane (i.e., perpendicular
  to the plane of the galactic disk.), exhibit a range of
  eccentricities sensitive to both the initial position and velocity
  of the GC. The potential of the dark matter halo and that of the
  central disk is included in all simulations.}
\begin{tabular}{cccc}
\hline \hline
Label & Initial Position & Initial Velocity  & Half-light Radius \\
      & (kpc)            & (kms$^{-1}$) & (pc)  \\ 
\hline
 GC-1 & (20.0, 0.0, 0.0) & (84.7, 0.0, 84.7) & 8 \\
 GC-2 & (35.0, 0.0, 0.0) & (84.7, 0.0, 84.7) & 8 \\
 GC-3 & (60.0, 0.0, 0.0) & (84.7, 0.0, 84.7) & 8 \\
 GC-4 & (35.0, 0.0, 0.0) & (63.0, 0.0, 63.0) & 8 \\
 GC-5 & (60.0, 0.0, 0.0) & (35.0, 0.0, 35.0) & 8 \\
 GC-6 & (35.0, 0.0, 0.0) & (84.7, 0.0, 84.7) & 3 \\
\hline
\end{tabular}
\label{tab:sims}
\end{table}

\begin{figure}
\centering
\center{\includegraphics[width=\columnwidth]{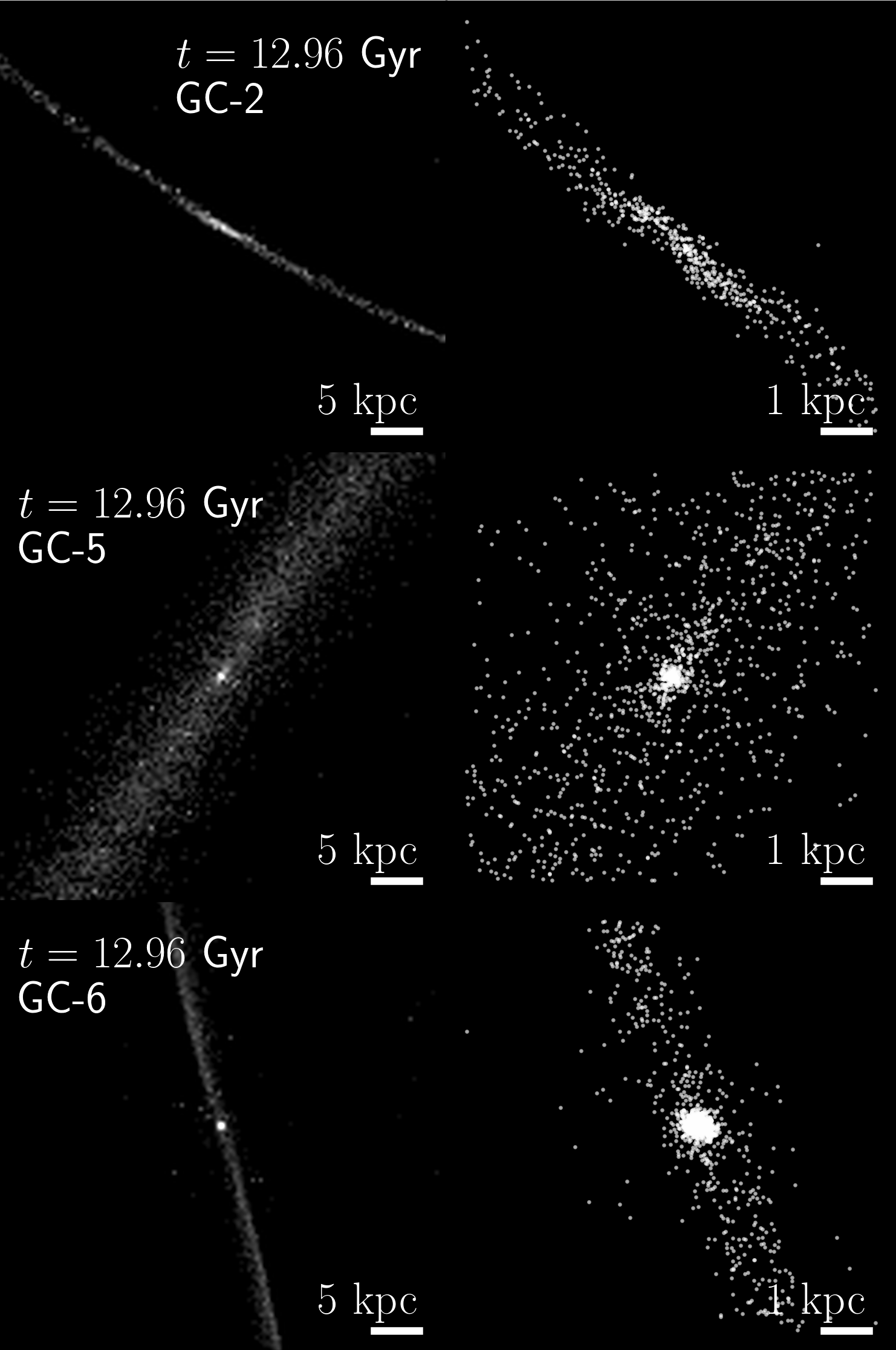}}
\figcaption{$z=0$ projections of GC streams in three of the
  simulations that we have performed. In each panel, the projection is
  centered on the cluster, displaying a region 50 kpc in size (left
  column), and a zoomed-in region 10 kpc in size (right column).}
\label{fig:GC_projections}
\end{figure}

Figure~\ref{fig:GC_projections} shows projections of a subset of our
simulations at the present time. As the initial GC structure is
identical in each run, different stream morphologies can be attributed
to the different choice of initial configurations as listed in
Table~\ref{tab:sims}. For example, while the cluster is initialized
with the same orbital parameters in GC-2 (top row) and GC-6 (bottom
row), the more compact initial structure of GC-6 ($r_h=3$ pc) compared
to GC-2 ($r_h=8$ pc) means that the former is less susceptible to
tidal disruption, resulting in a more prominent cluster core at
$z=0$. On the other hand, while GC-5 (middle row) shares identical
structural properties to GC-2, it also experiences less tidal
disruption for the simple reason that it undergoes fewer pericentric
passages than GC-2 \citep[see, e.g.][]{Penarrubia2009}.

\begin{figure*}
\center{\includegraphics[scale=0.35]{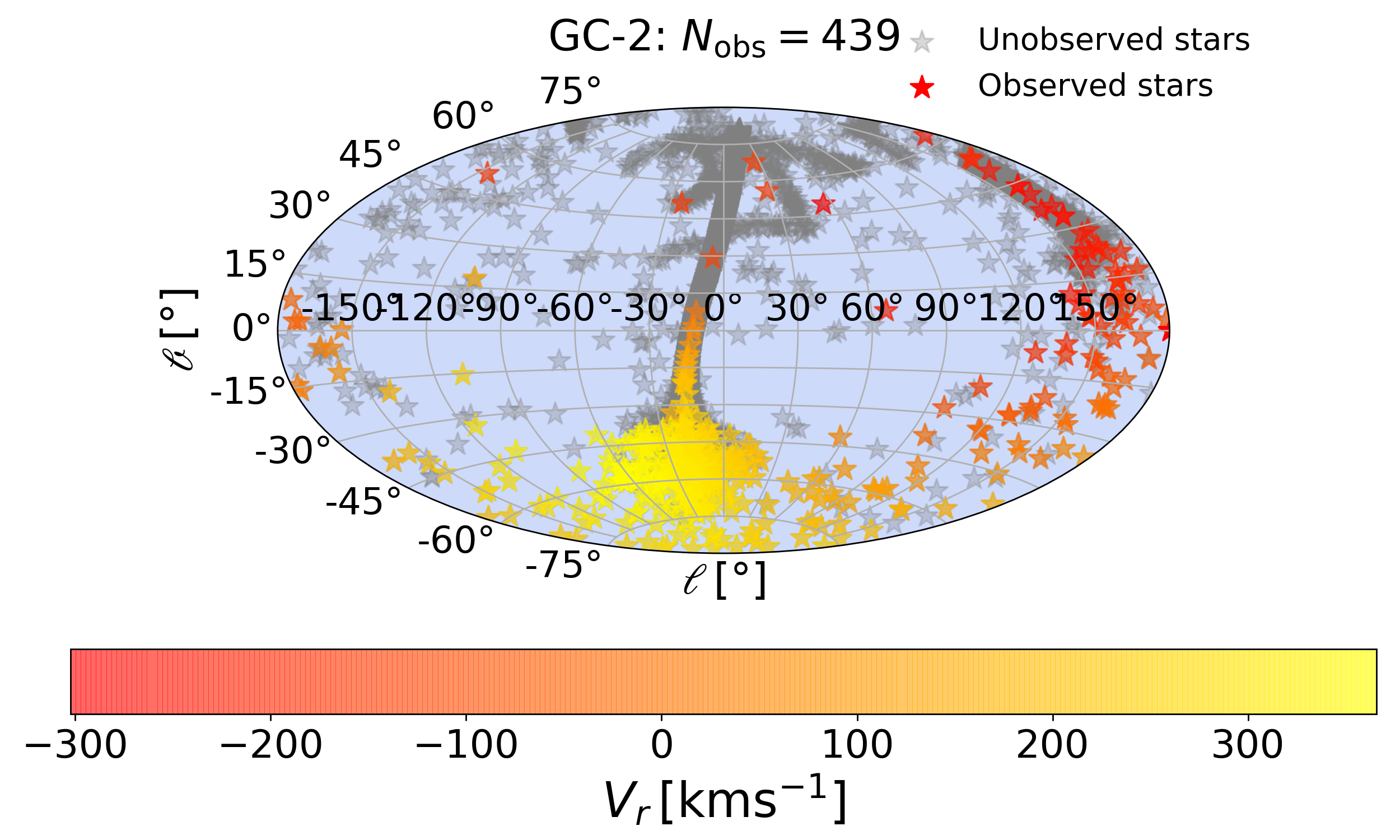}
  \includegraphics[scale=0.35]{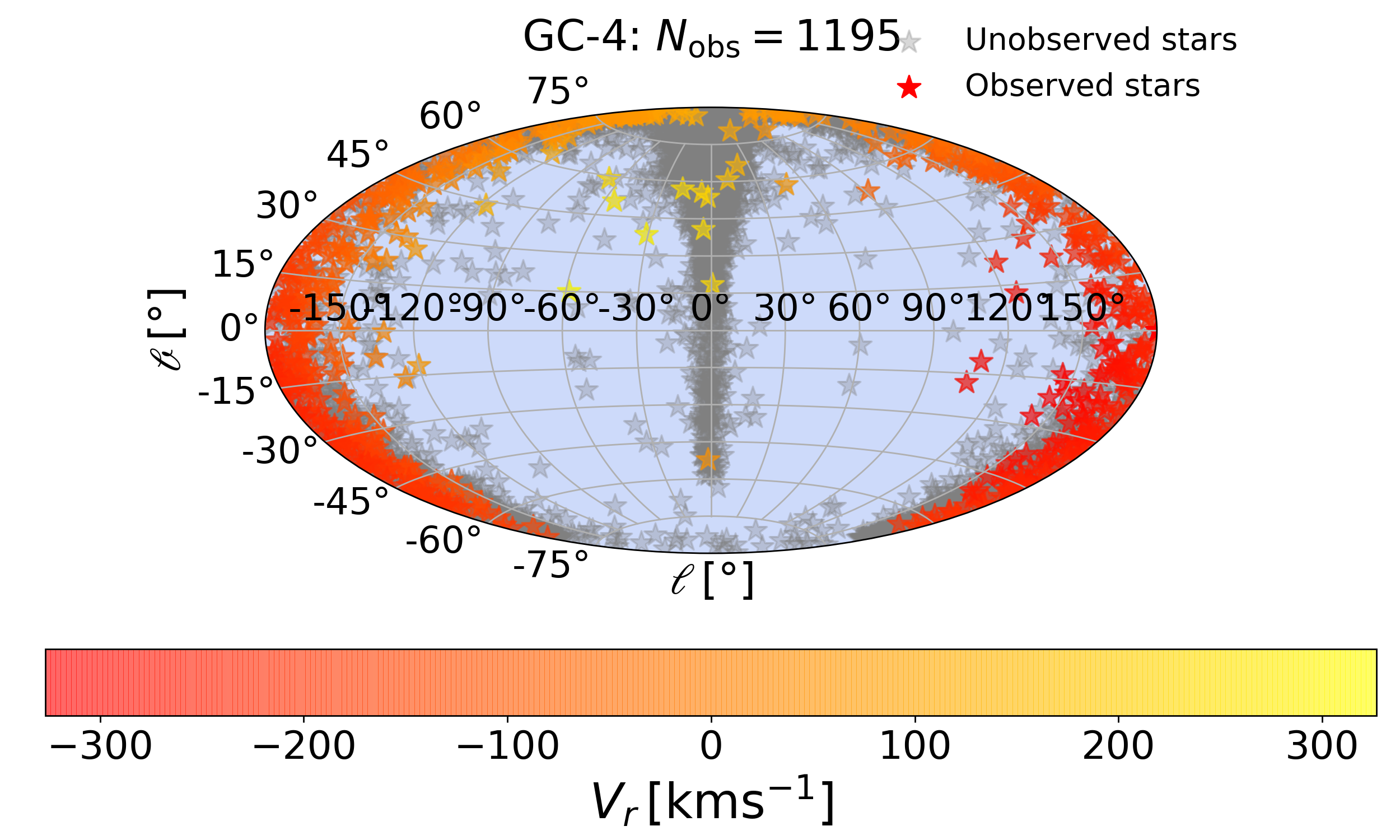}
  \\ \includegraphics[scale=0.35]{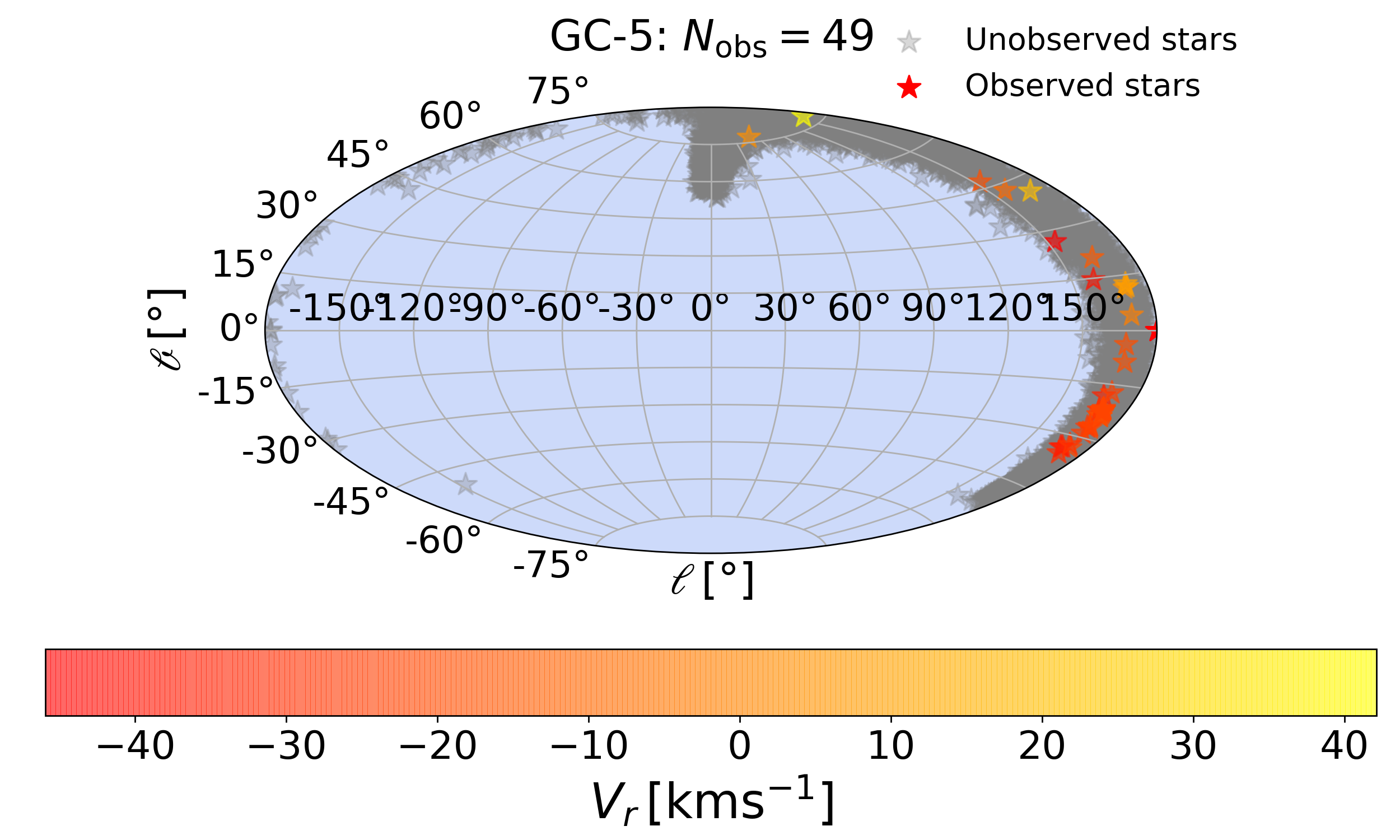}}
\figcaption{All-sky projections of streams at $z=0$. Stars that pass
  \gaia{}'s detection limit of $G \approx 20$~mag are shown in color;
  stars fainter than this limit are shown in gray. Individual colors
  denote the radial velocity of each star measured relative to an
  observer on the solar circle.}
\label{fig:Gaia_view}
\end{figure*}

\section{Simulated streams as observed by gaia} \label{sec:gaia}

In what follows, we emulate mock \gaia{} observations of our simulated
GC streams by projecting the stellar distribution onto all-sky
maps. We convert between cartesian coordinates used in the simulation
to heliocentric latitude, $\mathcal{b}$, and longitude, $\mathcal{l}$,
using the {\sc astropy} package \citep{astropy2018}. For each star, we
use its mass to calculate its luminosity, and use its distance from
the observer to compute the equivalent apparent magnitude,
$G$. \gaia{} is expected to be complete down to a nominal magnitude
limit of $G \approx 20$~mag; this threshold can be used to determine
which sections of each simulated stream would be within \gaia{}'s
detection capability.

Figure~\ref{fig:Gaia_view} displays all-sky projections of streams
from three of our simulations (GC-2, GC-4 and GC-5) as they would be
seen by an observer on the solar circle around the center of the Milky
Way. Stars that would be observed by \gaia{} are shown in color; the
range of colors corresponds to the radial velocity, $V_r$, of the star
relative to the observer. Unobserved portions of the stream (i.e.,
composed of stars fainter than $G=20$~mag) are shown in gray. In each
panel, we determine an `optimal' observer position as the point on the
solar circle that maximizes the number of observed stars.

As expected, the number of stars observed at present depends on the
initial orbital configuration for each simulation. For example, both
GC-2 and GC-4 are initialized 35 kpc from the center of the Milky Way,
but owing to its lower initial orbital velocity (see
Table~\ref{tab:sims}), GC-4's motion is more tightly bound by the disk
potential, ending up at a galactocentric distance of $\sim 12$~kpc at
$z=0$ (rather than $\sim 20$~kpc in the case of GC-2). The result is
that almost three times as many stars in GC-4 fall within \gaia{}'s
observable window compared to GC-2. Similarly, GC-5, which started off
60 kpc from the center of the Milky Way, displays the smallest number
of observed stars. In the remainder of this Letter, we will be
concerned only with the set of observed stars in each simulation.

\begin{figure*}
\center{\includegraphics[width=\columnwidth]{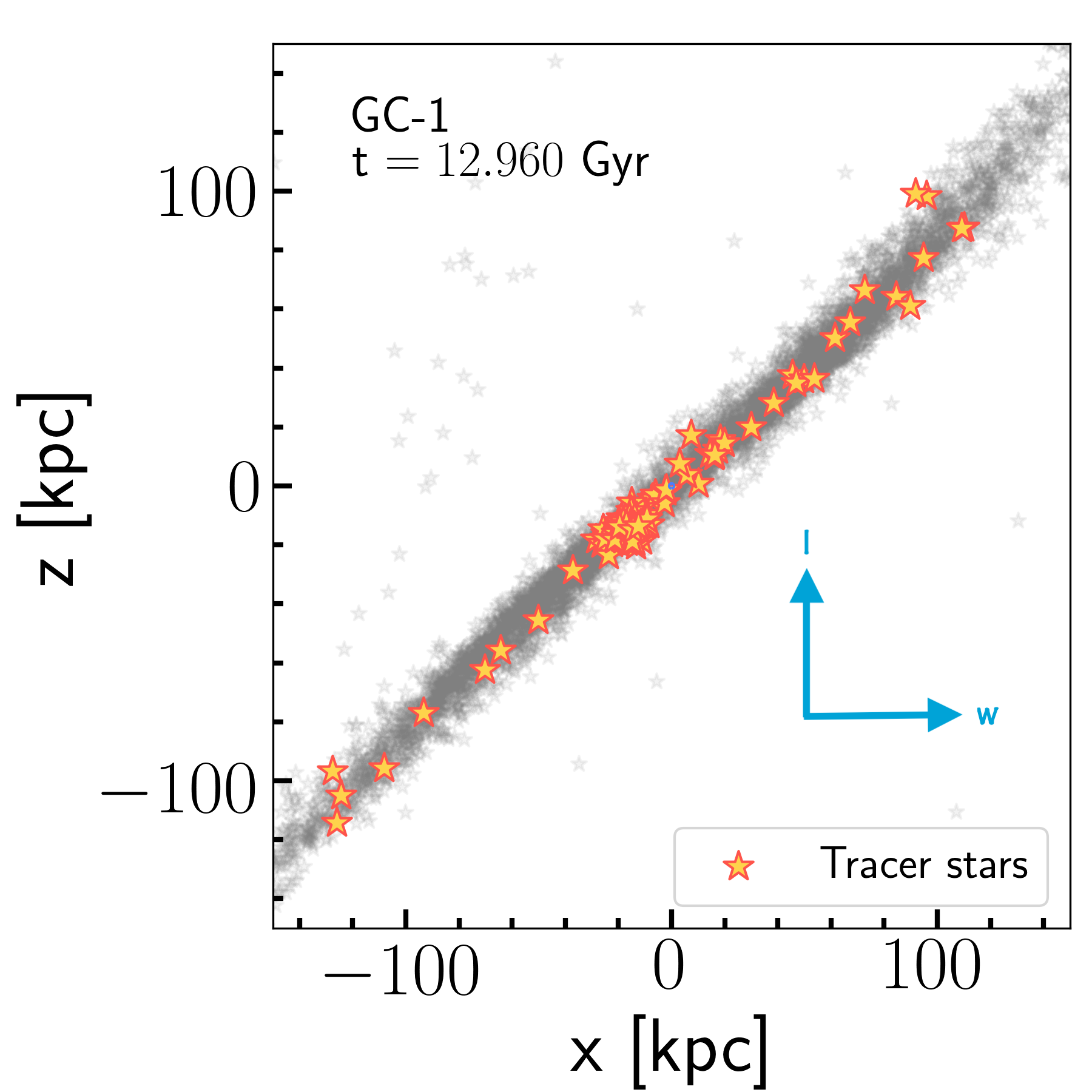}
  \\ \includegraphics[width=\columnwidth]{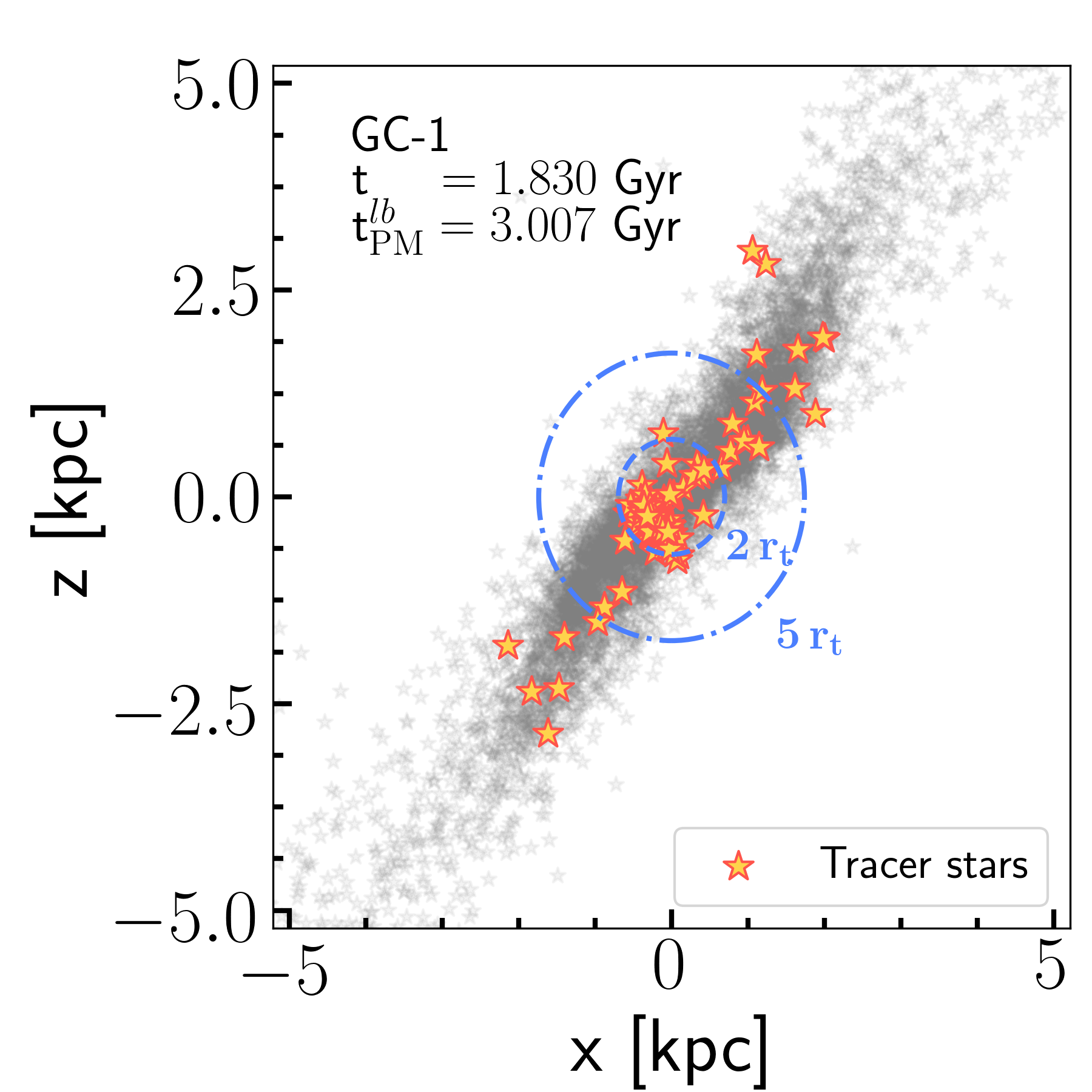}
  \includegraphics[width=\columnwidth]{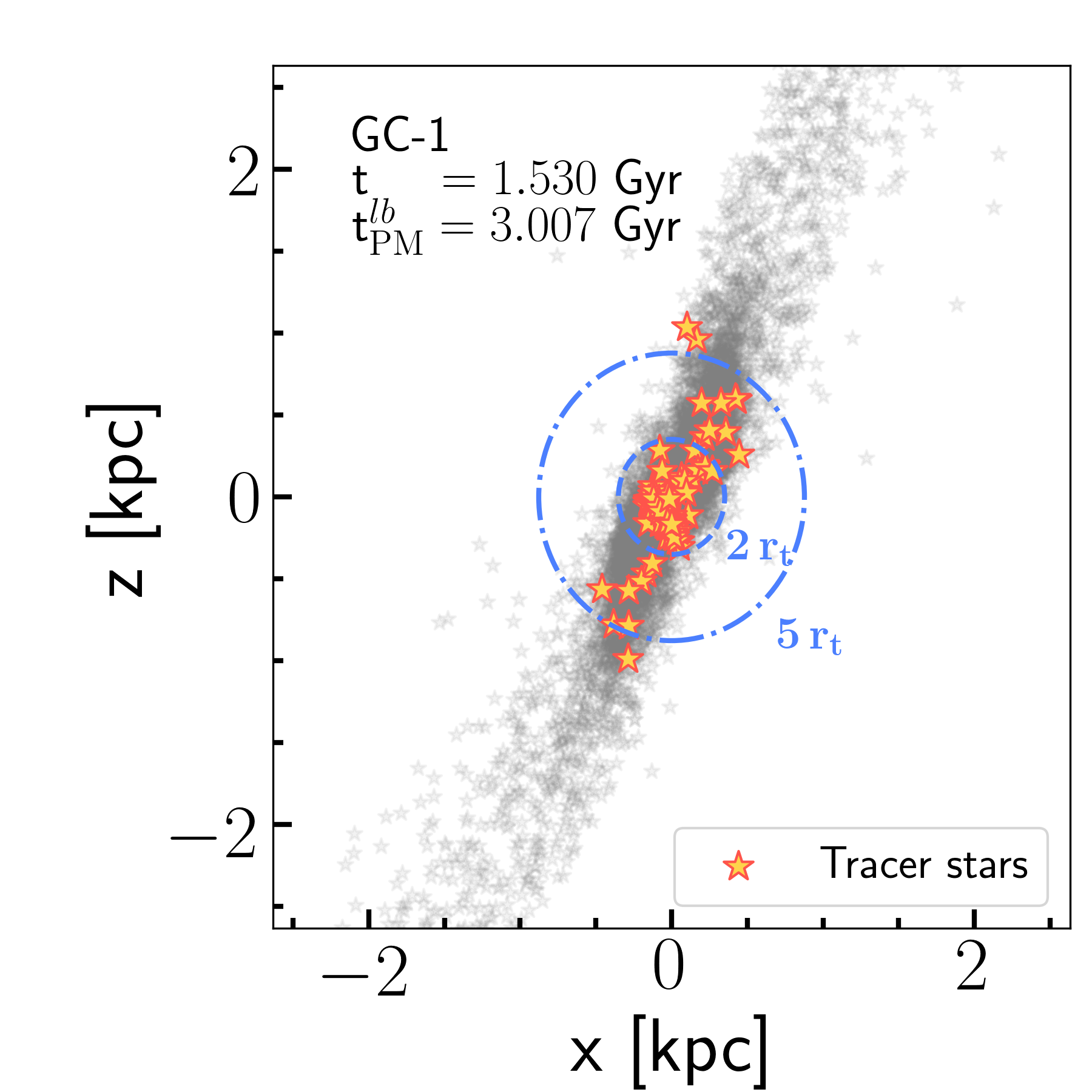}}
\figcaption{Tracking the time evolution of tracer stars (i.e., those
  that would be observed by \gaia{}; orange) in the GC-1
  simulation. After identifying these stars at the present time (top
  panel), we determine the time since the disruption of the stream as
  the epoch when {\it at least half} of all stars in the stream were
  contained within $5r_t$ ($t=1.83$ Gyr; bottom left panel) or $2r_t$
  ($t=1.53$ Gyr; bottom right panel). The lookback time estimated from
  the $z=0$ proper motions of the tracer stars (using
  Eq.~\ref{eq:tpm}), $t^{lb}_{{\rm PM}} = 3.01$ Gyr.}
\label{fig:Rewind}
\end{figure*}

\section{A characteristic timescale using proper motions} \label{sec:timescale}

Next, we demonstrate that a characteristic timescale that can be
defined using the proper motions of tracer stars in a stream matches
remarkably well the period that these stars have spent outside the
tidal influence of the GC from which they originated.

Given a set of stars observed in a stream at present day, we can label
their positions (velocities) in two orthogonal directions along the
plane of the sky as $X_w$ and $X_l$ ($V_w$ and $V_l$). The subscripts
on each quantity refer to the fact that they are measured along the
width and along the length of the stream -- corresponding to
measurements along the $x$ and $z$ axes, respectively, as shown in the
upper panel of Figure~\ref{fig:Rewind}. These quantities are directly
related to the parallax and proper motion in $\mathcal{l}$ and
$\mathcal{b}$ coordinates using the relations:
\begin{equation}
\begin{aligned}
X_w &= r_{\odot} - r_{{\rm obs}} \cos{\mathcal{b}} \cos{\mathcal{l}}, \\
X_l &= r_{{\rm obs}} \sin{\mathcal{b}},\\
V_w &= V_{\odot,x} + V_r \cos{\mathcal{b}} \cos{\mathcal{l}} \\ &- r_{{\rm obs}} \left[ \mu_{\mathcal{b}} \cos{\mathcal{l}} \sin{\mathcal{b}} + \mu_{\mathcal{l}} \cos{\mathcal{b}} \sin{\mathcal{l}} \right], \\
V_l &= V_{\odot,z} + V_r \sin\mathcal{b} + r_{{\rm obs}} \mu_{\mathcal{b}} \cos{\mathcal{b}},
\end{aligned}
\end{equation}
where $r_{{\rm obs}}$ is the radial distance to the star from the
observer, $\mu_{\mathcal{l}}$ and $\mu_{\mathcal{b}}$, respectively,
are the proper motions in the $\mathcal{l}$ and $\mathcal{b}$
directions, $r_\odot = 8.3$ kpc is the distance to the Galactic center
and $V_\odot = \left(11.1, 232.24, 7.25 \right)$ kms$^{-1}$ is the
solar motion relative to it \citep{Schonrich2010,Bovy2015}. For these
tracers, we can then define the dispersion in position, $\sigma_x$,
and velocity, $\sigma_v$, along the stream as:
\begin{equation} \label{eq:disp}
\begin{aligned}
\sigma_x &=& \sqrt[]{\sigma_{X_w}^2 + \sigma_{X_l}^2}\,, \\
\sigma_v &=& \sqrt[]{\sigma_{V_w}^2 + \sigma_{V_l}^2} \,, 
\end{aligned}
\end{equation}
Finally, we define a characteristic timescale ascertained from these
proper motions, $t_{{\rm PM}}$, given by:
\begin{equation} \label{eq:tpm}
t_{{\rm PM}} = \frac{\sigma_x}{\sigma_v}.
\end{equation}
Once stars have escaped the tidal influence of the GC, their motions
should be dominated by acceleration due to the external galactic
potential these stars are embedded in, rather than the GC itself. For
an external potential that evolves reasonably slowly in time, the
timescale provides an estimate for the duration of the period that the
escaped stars have been under the influence of the external
potential. Consequently, if the age of a stream is defined as the time
since the disruption of the GC resulting in the stream, the timescale
defined by Eq.~(\ref{eq:tpm}) should be correlated with the age of the
stream itself.

To investigate whether this is indeed the case, we followed the
evolution of the tracer stars backward in time in each simulation to
find out when their dynamics are no longer dominated by the GC. This
regime can be identified using the tidal radius, $r_t$, which sets the
radial distance from the center of the GC at which the potential of
the cluster is balanced by the background potential.  Theoretically,
we estimate the tidal radius as a function of redshift as:
\begin{equation}
r^3_t(z) \approx \frac{{\rm G}M_c(z)}{V_{\rm circ}^2(z)} r^2_{{\rm gc}}(z),
\end{equation}
\citep[c.f.,][]{Dehnen2004} where $V_{\rm circ}(z)$ is the circular
velocity of the host halo (measured at $r_{200}$) at redshift, $z$,
while $M_c(z)$ and $r_{{\rm gc}} (z)$, respectively, are the mass and
galactocentric distance of the star cluster at this redshift. Each of
these quantities are estimated at the output times of our
simulations. Over the course of the simulations, $V_{\rm circ}$
increases while $M_c$ decreases, resulting in an overall decrease in
$r_t$ as a function of time. We consider two possible criteria for a
star to be within the tidal influence of the cluster: (i) $r < 2r_t$
and (ii) $r < 5r_t$, where $r$ is the distance of a given star
particle from the center of the cluster.  More specifically, we
determine the age of the stream using the two tidal radii criteria,
$t_{{\rm tidal}}$, as the last epoch when {\it at least 50\%} of the
{\it all} stars in the stream (i.e., observed and unobserved) satisfy
conditions (i) or (ii). Condition (i) is more commonly adopted in the
literature \citep[e.g.][]{Wilkinson2003,Aarseth2012,Madrid2014};
additionally, considering the more restrictive condition (ii) gives a
handle of the uncertainty in measuring the time since tidal escape. By
comparing $t_{{\rm PM}}$ with $t_{{\rm tidal}}$, we can test how
accurately Eq.~(\ref{eq:tpm}) can be used to determine the time since
the disruption of the GC.

Figure~\ref{fig:Rewind} is an illustration of this comparison for the
GC-1 simulation. The stars in the stream marked in orange (158 in
total) are those that could be observed by \gaia{}, and act as the
tracer population for estimating the age of the stream. Using
Eqs.~(\ref{eq:disp}) \&~(\ref{eq:tpm}), we estimate the age to be
$9.95$ Gyr, corresponding to a lookback time of $t^{lb}_{{\rm PM}}
\approx 3.01$ Gyr. To estimate the ``dynamical'' age of the stream, we
then trace the evolution of the entire set of stars (tracers and
unobserved) through the GC-1 simulation, and find that at least half
of this population satisfies condition (i) at $t=1.53$ Gyr and
condition (ii) at $t=1.83$ Gyr.

\begin{figure}
\center{\includegraphics[width=\columnwidth]{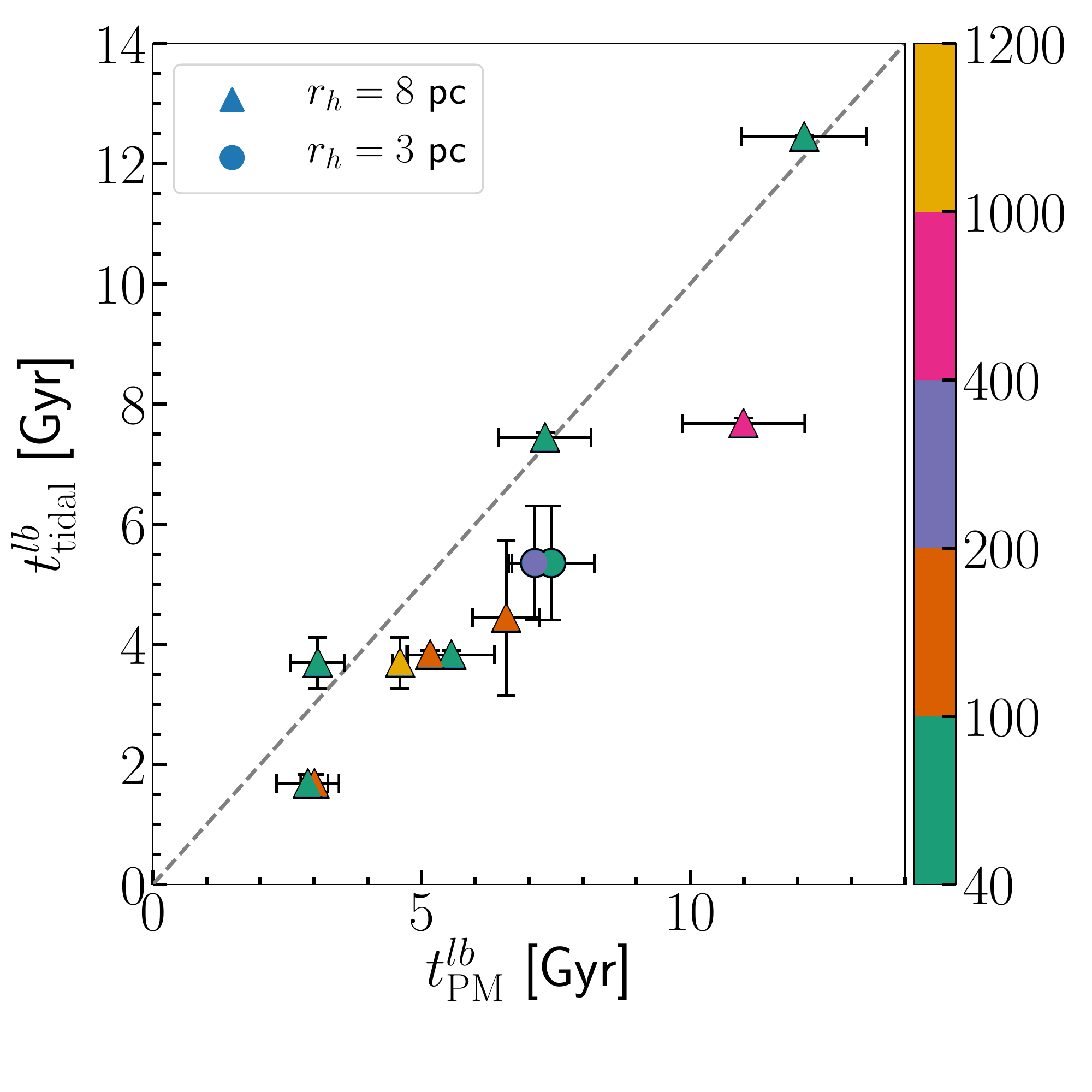}}
\figcaption{Relation between $t^{lb}_{{\rm PM}}$ (the age of a stream
  estimated using proper motions) and $t^{lb}_{{\rm tidal}}$ (the age
  inferred from tracing star particles backward in time) for the
  simulations performed in this work. The vertical error bars
  represent the difference in measuring $t^{lb}_{{\rm tidal}}$ using
  either $2r_t$ or $5r_t$ as the critical radius for escape. The
  horizontal error bars represent the uncertainty in estimating the
  age using Eq.~(\ref{eq:tpm}) after propagating through the typical
  errors with which proper motions and parallax would be measured by
  \gaia{} given the apparent magnitudes of the tracer stars in each
  simulation. The color scale on the right indicates the number of
  tracers used to estimate the age.}
\label{fig:Predictions}
\end{figure}

The result of this analysis for our other simulations is shown in
Figure~\ref{fig:Predictions}. We observe streams in each simulation at
two output times. In some cases, the same part of the stream is
observed twice; these correspond to data points that have the same
value of $t^{lb}_{{\rm tidal}}$. Given the apparent magnitude of each
tracer star in the simulated stream, we estimate errors on its proper
motion and parallax as they would be measured by \gaia{} using the
{\sc PyGaia}
package\footnote{\href{https://github.com/agabrown/PyGaia}{https://github.com/agabrown/PyGaia}};
these errors are combined to get a rough estimate of the error on
$t^{lb}_{{\rm PM}}$. The error on any given measurement therefore
depends on both the size of the tracer population (which we indicate
using the color scale of each data point in
Figure~\ref{fig:Predictions}) and the brightness of the stars in this
set (or, equivalently, the proximity of the stream). 

In general, the age of the stream estimated using proper motions
matches quite well the lookback at which these stars escaped the tidal
radius of the GC. The agreement between $t^{lb}_{{\rm PM}}$ and
$t^{lb}_{{\rm tidal}}$ is typically within 20-40\% in the majority of
the experiments that we have carried out. Encouragingly, this level of
agreement is also true for GC-6 ($r_h=3$ pc), for which the tidal
disruption rate (and, consequently, the distribution of energy and
angular momenta of the disrupted stars) is different to the other
examples that we have considered.

Figure~\ref{fig:Predictions} shows that $t^{lb}_{{\rm PM}}$
underpredicts the ``true'' age of the stream, $t^{lb}_{{\rm
    tidal}}$. The reason for this systematic difference can be
ascribed to the population of stars chosen to trace the age using
proper motions. Limiting the estimate of $t^{lb}_{{\rm PM}}$ to only
``observed'' stars preferentially selects more massive stream
members. Due to mass segregation within a GC, massive stars tend to
sink toward the cluster center while low-mass stars move farther away
-- these low-mass stars are the first to exit the tidal radius of the
cluster upon disruption. Using only the more massive, observed stars
slightly underestimates the actual time since disruption, and should
therefore be interpreted as a lower bound on the age of the stream.

It is worth highlighting that Eq.~(\ref{eq:tpm}) does not necessarily
reflect the age of the GC itself; instead, it is a measure of the time
at which the tidal disruption of the globular cluster resulted in a
given part of the observed stream. An observational determination of
this time can be used to constrain the assembly history of the Milky
Way.

\section{Conclusions} \label{sec:conclusions}

By virtue of their ancient stellar populations, GCs represent the most
interesting astrophysical entities found in galaxies. As these dense
concentrations of stars orbit their host galaxy, they are continually
stripped by the tidal field of their host, resulting in the formation
of cold, extended stellar streams.

The \gaia{} mission will provide precise measurements of the parallax
and proper motions for more than a billion stars in the Milky Way,
paving the way for a deeper understanding of the kinematics of stars
and the assembly history of our Galaxy. In this Letter we have shown
that, using a sample of tracer stars in a stellar stream, it is
possible to use the proper motions of these stars to infer the epoch
at which a globular cluster was disrupted, resulting in the formation
of the stream. Specifically, we find that a timescale defined using
the dispersion in positions and velocities of stars in the plane of
the stream (Eq.~\ref{eq:tpm}) provides a good estimator for how long
these stars have spent outside the tidal radius of the cluster.

We verified our results by running a sequence of simulations evolving
a cluster of mass $10^5\,M_\odot$ in a time-evolving potential
comprising the dark matter halo of the Milky Way (extracted from a
cosmological zoom simulation) and a (static) central disk
(Section~\ref{sec:numerical}). As shown in
Figure~\ref{fig:Predictions}, the age of the stream inferred using
astrometric information of tracer stars at $z=0$ matches very well the
epoch at which these stars were last contained within the tidal radius
of the globular cluster. The procedure outlined in this Letter can
therefore be used as an effective method for dating the tidal
disruption of a globular cluster, which in turn serves as a lower
bound on the age of the cluster itself.

In reality, correctly identifying stream members in the observed data
is more challenging than simply assuming a magnitude cut as we have
done in this work. In particular, misidentifying stream members can
result in incorrectly measured dispersions, which would subsequently
propagate as a large error in the inferred age of the stream. Radial
velocities, where available, can be useful: GC streams are typically
very cold, and ``true'' stream members are likely to be clustered in a
phase diagram of radial distance and line-of-sight velocity. Jointly
considering both the proper motions and color-magnitude diagram can
also be informative in faithfully separating members of a stellar
stream from foreground/background contaminants. The \gaia{} DR2
dataset will be particularly transformative in this regard.

\acknowledgments

We are thankful to the referee and the editor for making a number of
suggestions that have significantly improved the overall quality of
this manuscript.  We are grateful to Long Wang for providing extensive
support in the setup and use of \nbody{}, and to Adrian Jenkins for
giving us access to his codes for generating cosmological initial
conditions. We benefited from some useful conversations with Ana
Bonaca.  We are thankful to the developers of {\sc astropy} and {\sc
  PyGaia} for maintaining these codes and making them public. S.B. is
supported by Harvard University through the ITC Fellowship. S.B. also
acknowledges the hospitality provided by the Kavli Institute for
Theoretical Physics in Santa Barbara during the ``Small-Scale
Structure of Cold(?)  Dark Matter'' programme, where part of this work
was completed. This research was supported in part by the National
Science Foundation under grant No. NSF PHY-1748958, and in part by the
Black Hole Initiative, which is funded by a grant from the John
Templeton Foundation.


\end{document}